\documentclass[aps,pra,reprint,superscriptaddress,nofootinbib,longbibliography]{revtex4-2}

\usepackage[unicode=true,
 bookmarks=true,bookmarksnumbered=false,bookmarksopen=false,
 breaklinks=false,pdfborder={0 0 1},backref=false,colorlinks=true]
 {hyperref}
 \hypersetup{
 linkcolor=magenta, urlcolor=blue, citecolor=blue, pdfstartview={FitH}, unicode=true}

\usepackage[utf8]{inputenc}

\usepackage{amsmath}
\usepackage{amssymb}
\usepackage{amsfonts}
\usepackage{graphicx} 
\usepackage{hyperref} 
\usepackage{xcolor}
\usepackage[normalem]{ulem}
\usepackage{placeins}

\begin{document}

\title{Memory-induced long-range order drag}

\author{Yuan-Hang Zhang}
\email{yuz092@ucsd.edu}
\thanks{Equal contributions.}
\affiliation{Department of Physics, University of California San Diego, La Jolla, CA 92093}

\author{Chesson Sipling}
\email{csipling@ucsd.edu}
\thanks{Equal contributions.}
\affiliation{Department of Physics, University of California San Diego, La Jolla, CA 92093}

\author{Massimiliano Di Ventra}
\email{email: diventra@physics.ucsd.edu}
\affiliation{Department of Physics, University of California San Diego, La Jolla, CA 92093}

\begin{abstract}
Recent research has shown that memory, in the form of slow degrees of freedom, can induce a phase of long-range order (LRO) in locally-coupled fast degrees of freedom, producing power-law distributions of avalanches. In fact, such memory-induced LRO (MILRO) arises in a wide range of physical systems. Here, we show that MILRO can be transferred to coupled systems that have no memory of their own. As an example, we consider a stack of layers of spins with local feedforward couplings: only the first layer contains memory, while downstream layers are memory-free and locally interacting. Analytical arguments and simulations reveal that MILRO can indeed drag across the layers, enabling downstream layers to sustain intra-layer LRO despite having neither memory nor long-range interactions. This establishes a simple, yet generic mechanism for propagating collective activity through media without fine tuning to criticality, with testable implications for neuromorphic systems and laminar information flow in the brain cortex.

\end{abstract}

\maketitle

\section{Introduction}

Long-range order (LRO) is typically associated with systems tuned to a critical point, where correlations span the entire system and collective dynamics emerge~\cite{tauber2014critical, pruessner2012self, hinrichsen2000non}. Recent work, however, has shown that LRO can also occur away from criticality when memory—in the form of slow, auxiliary degrees of freedom—interacts with fast, locally coupled variables~\cite{sipling2025memory}. This memory-induced LRO (MILRO) produces robust power-law correlations in the fast degrees of freedom, and has been observed in diverse contexts including spin-glass models~\cite{sipling2025memory}, quantum spin chains~\cite{weber2022dissipation}, neuromorphic devices~\cite{zhang2024collective}, models of neural activity~\cite{sun2024memory}, memcomputing machines~\cite{di2022memcomputing,sipling2025phase}, and feedback-induced quantum phase transitions~\cite{ivanov2020feedback}. In these systems, slow memory variables act as an effective long-range interaction {\it in time}, inducing spatial LRO in the fast degrees of freedom over a wide range of parameters, and offering a mechanism for collective dynamics that does not require fine parameter tuning.

It is worth noticing that in all the above examples, the memory degrees of freedom share the same spatial region as the fast degrees of freedom. An open question is then whether such MILRO can propagate even beyond spatial regions of the system where the memory resides. Answering this question is not just an academic exercise. Rather, it is motivated by the fact that in many natural and engineered systems, functional units are arranged in modules or layers that interact through local couplings. A prototypical example is the brain architecture, where the neocortex is organized into distinct layers with structured feedforward and feedback connectivity, enabling hierarchical and laminar information processing~\cite{harris2015neocortical, douglas2004neuronal, mejias2016feedforward, felleman1991distributed}. If a subset of these layers contains memory-rich dynamics, hence possibly supporting MILRO, can their order ``drag'' downstream into memory-free layers, allowing the entire stack to sustain collective, scale-free activity?\footnote{Furthermore, could this effect explain the experimentally observed scale-free dynamics in the firing of neurons~\cite{beggs2003neuronal}?}

Here, we address these questions by considering a stack of spin-glass layers with unidirectional, local feedforward couplings; see Fig.~\ref{fig:main_fig_1}. Only the base layer contains slow memory variables; all other layers evolve via memory-free spin-glass dynamics with short-range in-plane interactions. We find that, for a wide range of parameters, the MILRO in the driving layer propagates to deeper layers, which themselves exhibit robust intra-layer LRO despite lacking memory and long-range couplings.

We term this phenomenon (memory-induced) {\it long-range order drag} in analogy with the Coulomb drag~\cite{narozhny2016coulomb, jauho1993coulomb, li2016negative}—where momentum transfer between nearby electron layers induces a current in the ``passive'' layer—and the spin drag, where spin currents in one spin species induce currents in another via some interactions~\cite{d2000theory, weber2005observation}. In our case, the ``dragged'' quantity is not a current but {\it an entire ordered state}, transmitted through local couplings.\footnote{We point out that we expect such a phenomenon to occur even if the origin of LRO in the base layer is not due to memory, and such LRO ``seed'' is not destroyed by the interaction with the other layers.} 

This mechanism suggests a general route for propagating collective dynamics across layered media without fine tuning to criticality. Beyond its theoretical interest, LRO drag may offer design principles for multi-layer neuromorphic architectures~\cite{li2017three, strukov2009four, kim20242d} and provide a potential explanation for sustained inter-laminar correlations in cortical circuits~\cite{harris2015neocortical, douglas2004neuronal, mejias2016feedforward, felleman1991distributed}.

\section{Analytical understanding of the MILRO drag}\label{sec:analysis}
\begin{figure}[t]
    \centering
    \includegraphics[width=0.8\columnwidth]{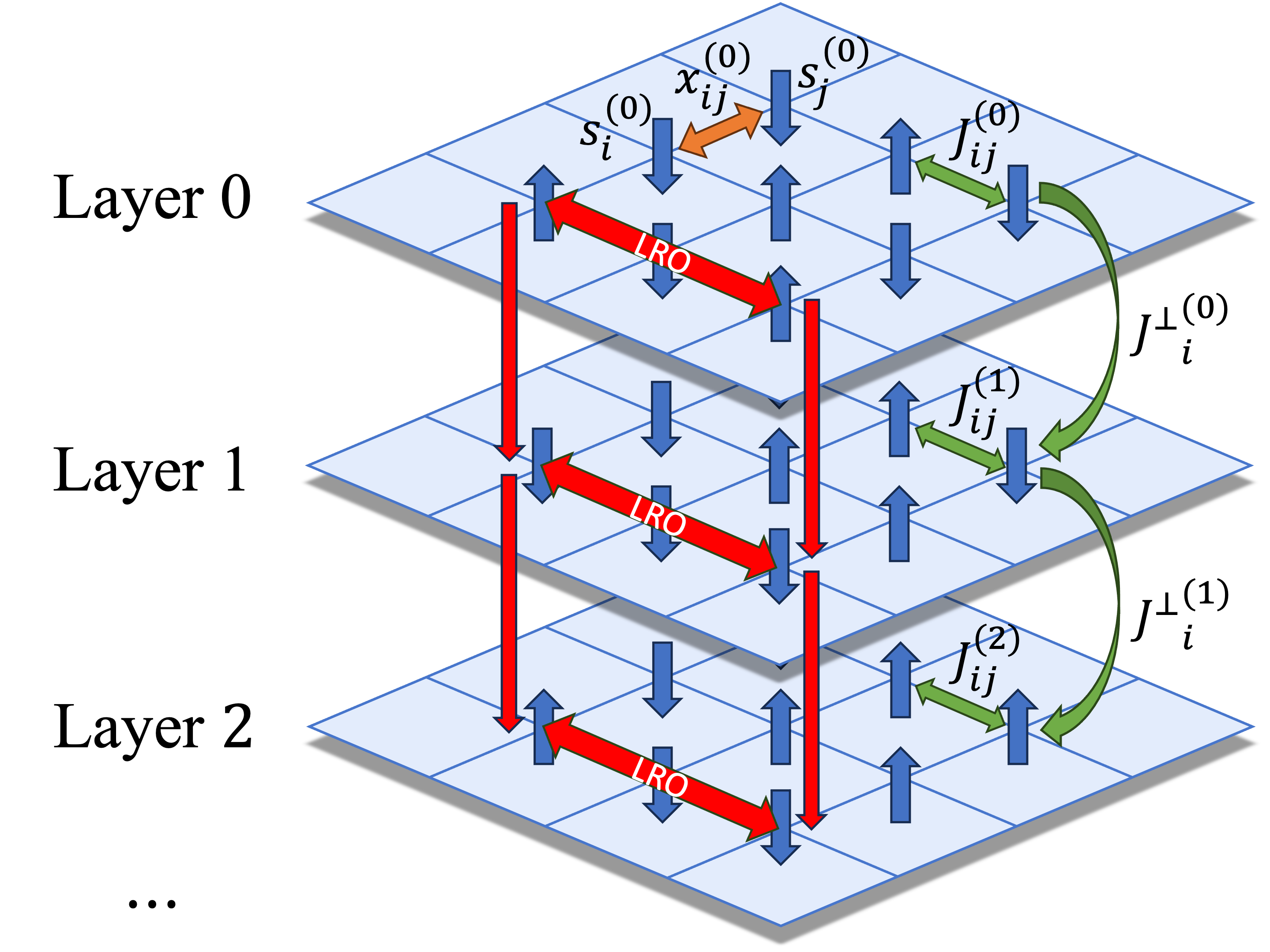}
    \caption{Schematic of the memory-induced LRO drag phenomenon in our system. In all layers, continuously relaxed spins interact via traditional spin-glass interactions. In the base $k = 0$ layer, spins also experience memory interactions, where the memory $x_{ij}$ is an auxiliary dynamical field which depends on the state of neighboring spins $s_i^{(0)}$ and $s_j^{(0)}$ and their mutual interaction $J_{ij}^{(0)}$. In deeper layers, memory does not appear, but instead, unidirectional, inter-layer interactions exist which enable the memory-induced LRO in layer 0 to be ``dragged'' through deeper layers.}
    \label{fig:main_fig_1}
\end{figure}

As anticipated, in order to study this phenomenon within the simplest system possible, we consider the model system depicted in  Fig.~\ref{fig:main_fig_1}. Each layer contains continuously relaxed spins with either ferromagnetic or anti-ferromagnetic interactions with their intra-layer neighbors. In the base layer, an additional memory degree of freedom (DOF) exists. Each memory DOF is defined on a lattice edge and tracks the recent history of its corresponding pair of spins; namely, it slowly grows or decays based on the degree of satisfaction of the relevant spin-glass interaction. In all other layers, these memories are not present. Instead, a unidirectional spin-glass interaction between neighboring {\it inter}-layer spins does exist, which enables layers that are closer to the base to dynamically influence deeper ones.\footnote{For simplicity, in this work we consider only a {\it feedforward} inter-layer interaction, inspired by feedforward neural networks~\cite{bebis2002feed}. However, we expect such a phenomenon to also occur if feedback connectivity is present, provided this feedback does not destroy the base MILRO.}

We note that this model is an extension of the one used in Ref.~\cite{sipling2025memory}, which considers only a single layer of spins with memories. That study acts as an essential background for the current work.

\subsection{Model dynamics}

Mathematically, the above system can be modeled by the following dynamical equations, which are an extension of the ones used in 
Ref.~\cite{sipling2025memory}:

\begin{equation}
\begin{aligned}
    \dot{s}_i^{(0)}=& \sum_{\langle ij \rangle} J_{ij}^{(0)}s_j^{(0)}-g\sum_{\langle ij\rangle}x_{ij}s_i^{(0)},\\
    \dot{s}_i^{(k)}=& \sum_{\langle ij \rangle} J_{ij}^{(k)}s_j^{(k)} + J_i^{\perp(k)}s_{i}^{(k-1)} \qquad (k\ge 1),\\
    \dot{x}_{ij} =& \gamma(C_{ij}-\delta), \quad C_{ij}=(J_{ij}^{(0)}s_i^{(0)} s_j^{(0)} + 1) / 2.\label{eq:model}
\end{aligned}
\end{equation}

Here, $s_i^{(k)}$ denotes the value of the spin variable at layer $k$, planar coordinate $i$.  Each 2D layer $k \geq 0$ consists of the standard, nearest-neighbor spin-glass interactions $J_{ij}^{(k)} = \pm J$, uniformly randomized in each layer; $\sum_{\langle ij \rangle}$ denotes summation over 2D nearest-neighbors. In the base layer $k = 0$ {\it only} do we introduce an additional degree of freedom (DOF) $x_{ij}$ which grows or decays based on the state of its two accompanying spins, $s_i^{(0)}$ and $s_j^{(0)}$. In all other layers $k \geq 1$, there is an additional unidirectional ferromagnetic/anti-ferromagnetic interaction $J_i^{\perp(k)} = \pm J^{\perp}$ (again, assigned randomly) which couples $s_i^{(k)}$ to $s_i^{(k-1)}$. We clamp these dynamics so that $s_i^{(k)} \in [-1, 1]$ and $x_{ij} \in [0, 1]$. As in Ref.~\cite{sipling2025memory} and unless otherwise specified, we fix $\delta = 3/4$, $g=2$ and $J = 1$ for the remainder of this work; $\gamma$ and $J^\perp$ are tunable parameters that we will study in the following sections. All summations are provided explicitly in these dynamical equations and following analytics; Einstein notation is used nowhere. A schematic representing this model can be found in Fig.~\ref{fig:main_fig_1}.

\subsection{MILRO Drag}\label{sec:MILROa}

Crucially, in the regime where $\gamma$ is small relative to the other characteristic frequencies of the system, $x_{ij}$ acts as a slowly-varying memory DOF which tracks the recent history of $s_i^{(0)}$ and $s_j^{(0)}$. It is precisely this memory DOF which has already been shown to induce LRO in the base layer $k = 0$ considered as a stand-alone layer~\cite{sipling2025memory}. Now, let us apply the same iterative integration approach used in~\cite{sipling2025memory} to Eqn.~\ref{eq:model}. We will work in the limit where the frequency of the memory DOFs, $\gamma$, is much smaller than all other characteristic frequencies in the problem (i.e., $\gamma << g$, $\gamma << J$, and $\gamma << J^\perp$) to infer how, under the appropriate conditions, long-range {\it intra}-layer correlations between high-$k$ spins can be induced.

First, we reiterate the results of the previous work~\cite{sipling2025memory} establishing MILRO in a single spin-glass layer with memory. The iterative integration scheme can be performed on the first layer in isolation (due to the unidirectional nature of the inter-layer interactions), on both the memory DOFs (from Eqn.~\ref{eq:model}):

\begin{equation}
\begin{split}
    x_{ij}(1/\gamma) &\approx x_{ij}(0) + \int_0^{1/\gamma} dt \, \gamma (C_{ij} - \delta), \\
    &\approx x_{ij}(0) + (J_{ij}^{(0)} \overline{s_{i, 0}^{(0)}} \overline{s_{j, 0}^{(0)}} + 1)/2 - \delta,
\end{split}
\end{equation}

\noindent and the spins:

\begin{equation}
\begin{split}
    s_i^{(0)}(1/\gamma) &\approx s_i^{(0)}(0) + \int_0^{1/\gamma} dt \, \bigg( \sum_{\langle ij \rangle} J_{ij}^{(0)} s_j^{(0)} -g \sum_{\langle ij \rangle} x_{ij} s_i^{(0)} \bigg), \\
    &\approx s_i^{(0)}(0) - \frac{g}{\gamma} \sum_{\langle ij \rangle} x_{ij}(0) \overline{s_{i, 0}^{(0)}} + \dots, \\
    s_i^{(0)}(2/\gamma) &\approx s_i^{(0)}(1/\gamma) - \frac{g}{\gamma} \sum_{\langle ij \rangle} x_{ij}(1/\gamma) \overline{s_{i, 1}^{(0)}} + \dots, \\
    &\approx -\frac{g}{2 \gamma} \overline{s_{i, 0}^{(1)}}  \sum_{\langle ij \rangle} J_{ij}^{(0)} \overline{s_{i, 0}^{(0)}} \overline{s_{j, 0}^{(0)}} + \dots,
\end{split}
\end{equation}
where $\overline{f}$ denotes the average of some quantity $f$ over the interval $t \in [l/\gamma, (l + 1)/\gamma)$. As this process is iterated further, additional, strong couplings begin to emerge between spins farther and farther from each other in the lattice. This induces an {\it effective} long-range interaction which emerges due to the presence of memory alongside the explicitly local spin-glass interactions~\cite{sipling2025memory}.

Next, let us consider how $s_i^{(k)}$ evolves in time for $k \geq 1$, concentrating on its correlation to spins in other layers at the same lattice position $i$. After a duration of time $T = 1/\gamma$, we can approximate $s_i^{(k)}(1/\gamma)$ as follows:

\begin{equation}
\begin{split}
    s_i^{(k)}(1/\gamma) &= s_i^{(k)}(0) + \int_0^{1/\gamma} dt \, \bigg( \sum_{\langle ij \rangle} J_{ij}^{(k)} s_j^{(k)} + J_i^{\perp(k)} s_i^{(k - 1)} \bigg), \\
    &\approx s_i^{(k)}(0) + \frac{1}{\gamma} \bigg( \sum_{\langle ij \rangle} J_{ij}^{(k)} \overline{s_{j, 0}^{(k)}} + J_i^{\perp(k)} \overline{s_{i, 0}^{(k - 1)}} \bigg).
\end{split}
\end{equation}

Importantly, when $J^\perp$ is sufficiently strong, the unidirectional inter-layer interactions will begin to dominate over the intra-layer ones. The necessary strength of the $J^\perp$ interaction which yields strong, inter-layer correlations will depend heavily on the degree of frustration within each spin-glass layer. In the most extreme case, where all of a spin's intra-layer neighbors are in energetic agreement with the corresponding spin-glass interaction $J_{ij}^{(k)}$ (i.e., where $J_{ij}^{(k)} s_j^{(k)} = 1$ for all nearest-neighbors $j$ of $i$), we must have $J^\perp/J > 4$ for a flip in layer $k-1$ to induce a flip at the corresponding site in layer $k$. Only then can the one inter-layer interaction overpower the four nearest-neighbor intra-layer interactions. However, in general, not all spins will necessarily align in this manner. We could also have a net of two spins being energetically aligned (the other two canceling each other out), a net zero alignment (complete cancellation), a net of two anti-aligning, or all four anti-aligning. 

Since $J_{ij}^{(k)} = \pm J$ are themselves chosen in a uniformly random manner (and because the spins spend most of their time at extremal values $s_i^{(k)} = \pm 1$; i.e., they flip rapidly relative to other dynamical timescales), let us assume the four nearest-neighbor $J_{ij}^{(k)} s_j^{(k)}$'s are drawn from a uniform distribution $\{-J, J\}$. In this case, the {\it strength} of the net intra-layer interaction $|J_{ij}^{(k)} s_j^{(k)}|$ takes values $\{0, 2J, 4J\}$ with probabilities $\{6/16, 8/16, 2/16\}$. This gives a mean strength of $\langle\Big|\sum_{ij} J_{ij}^{(k)} s_i^{(k)}\Big|\rangle\,\, =\, 3J/2$. Thus, $J^\perp/J > 3/2$ may be a more reasonable lower bound beyond which inter-layer interactions dominate at a majority of sites, if $s_i^{(k-1)} = \pm 1$ as well.

Of course, the exact lower bound of $J^\perp$ over which inter-layer correlations can easily spread to deeper layers, {\it propagating the entire collective state}, should be slightly higher than this value. That is because we anticipate some supermajority ($\gg 50\%$) of sites must be dominated by the inter-layer interaction in order for {\it system-wide} avalanches to propagate between layers; otherwise, the maximum number of spins which can participate in an avalanche will be capped far below the system size $L^2$. This will also depend on the system's dynamics and the precise distribution of $J_{ij}^{(k)}$ in each lattice. We will indeed confirm that this is the case in our numerical experiments reported in Sec.~\ref{sec:numerics}.

Regardless, it is clear that when $J^\perp$ is sufficiently large, we expect that $s_i^{(k)}$ and $s_i^{(k')}$ will begin to correlate strongly after some transient duration, even at moderate inter-layer distances $|k - k'| > 1$. Therefore, after a sufficiently large  number $N$ of timesteps of duration $1/\gamma$, when the inter-layer interaction is strong enough, a spin $s_i^{(k)}(N/\gamma)$ will correlate strongly with the average value of the spin $s_i^{(k - 1)}(t)$ in the interval $t \in [0, N/\gamma)$:

\begin{equation}
\begin{split}
    s_i^{(k)}(N/\gamma) &\approx \frac{J_i^{\perp(k)}}{\gamma} \bigg( \overline{s_{i, N-1}^{(k - 1)}} + \overline{s_{i, N-2}^{(k - 1)}} + \dots + \overline{s_{i, 0}^{(k - 1)}} \bigg) + \dots \\
    &\approx \frac{J_i^{\perp(k)}}{\gamma} \overline{s_i^{(k - 1)}}.
\end{split}
\end{equation}

Iterating this reasoning, we can infer that $s_i^{(k)}$ should {\it also} correlate with the corresponding spin in the {\it base} layer $s_i^{(0)}$, albeit more weakly than with the spin $s_i^{(k - 1)}$ in the adjacent layer.

Now, if this result is combined with the fact that the memory interactions induce long-range order in the base $k=0$ layer, it is now possible for the spins in layers $k \geq 1$ to correlate with each other over long distances due to their mutual correlation with the base layer. Essentially, after MILRO is established within spins in the base layer, the strong inter-layer coupling ``drags'' this collective state to deeper and deeper layers. Altogether, this constitutes a {\it multi-step} correlation process whereby LRO is transferred from a base layer {\it with} memory to additional layers {\it without} memory.

\section{Numerical verification of the MILRO drag}
\label{sec:numerics}

We now demonstrate the existence of MILRO drag numerically using avalanche statistics, supporting the previous theoretical analysis. Avalanches are defined as spatiotemporal bursts of spin flips separated by quiescent intervals; the size $s$ is the total number of flips in a burst. More numerical details are reported in Appendices~\ref{appendix:numerics} and~\ref{appendix:window}.

\subsection{Two-layer test and finite-size scaling}

\begin{figure*}[t]
    \centering
    \includegraphics[width=0.8\linewidth]{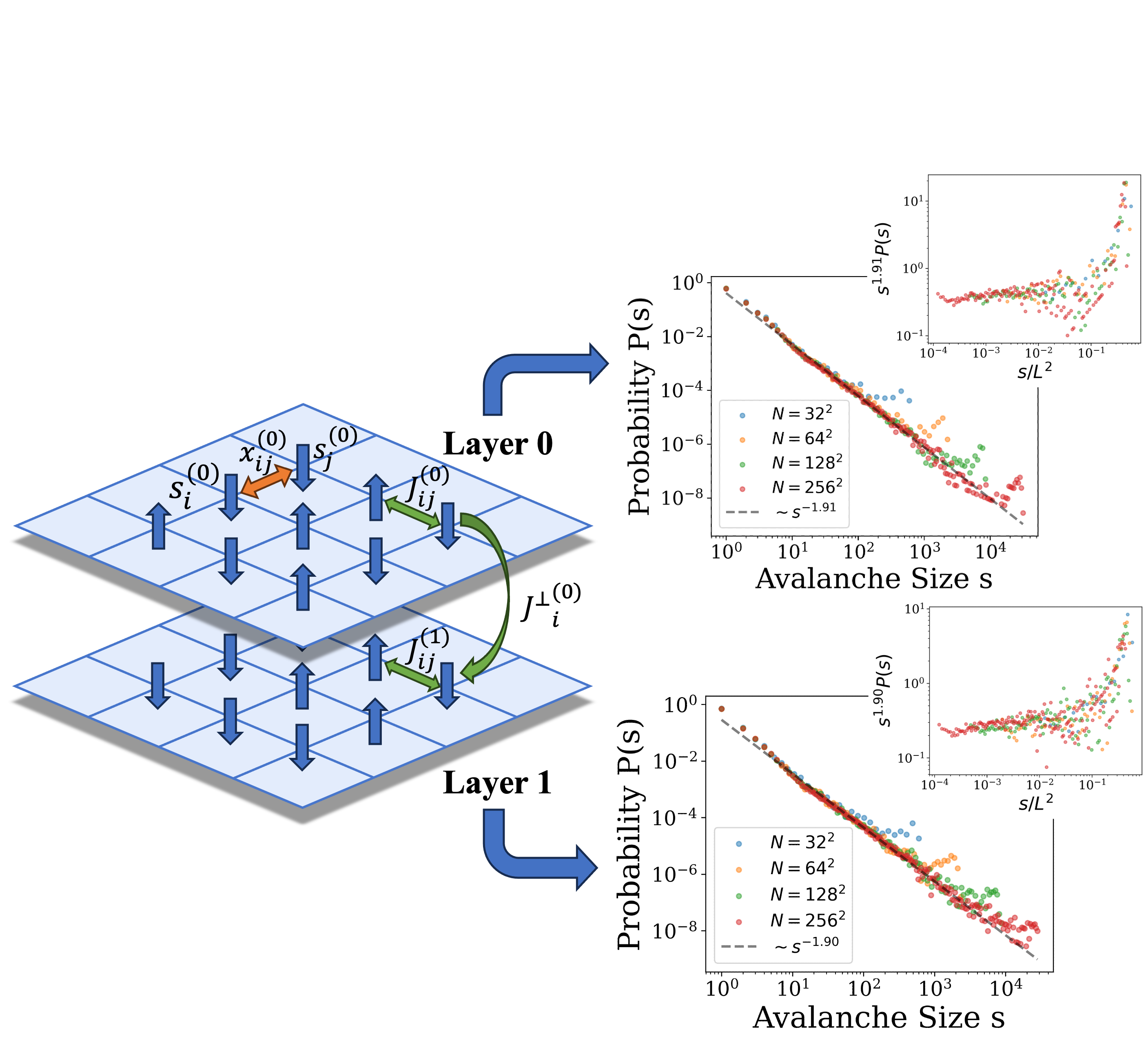}
    \caption{Schematic of two layers of continuously relaxed spins alongside avalanche size $s$ probability distributions $P(s)$. We notice that, in both layers, scale-free avalanche distributions exist which collapse well under finite-size scaling (shown in the insets). Furthermore, the scaling exponents in these distributions are nearly identical. This suggests that the collective, LRO state is effectively transferred from the base layer $k=0$ to the $k=1$ layer due to the inter-layer interactions. Both layers are simulated over 100 instances for $T=200$ simulation time (a.u.), with $\gamma = 0.2$ and $J^\perp = 3.5$.}
    \label{fig:finite_size_scaling}
\end{figure*}

We first simulate Eqn.~\ref{eq:model} on square lattices of size $L=\{32,64,128,256\}$ for the driving layer ($k=0$) and the first downstream layer ($k=1$); see Fig.~\ref{fig:finite_size_scaling}. The distributions $P_k(s)$ of avalanche size $s$ exhibit clear power laws over several decades, with finite-size characteristic bumps at large $s \lesssim L^2$. Assuming scale invariance,
\begin{equation}
    P_k(s;L)= s^{-\tau_k}\, f_k\!\left(\frac{s}{L^{D}}\right),
    \label{eq:finite_size}
\end{equation}
should hold~\cite{tauber2014critical}. Using the $\tau_k$ extracted from Fig.~\ref{fig:finite_size_scaling}, we do indeed obtain excellent collapses of $s^{\tau_k}P_k(s)$ versus $s/L^{D}$ with $D= 2$ for both $k=0$ and $k=1$ (insets of Fig.~\ref{fig:finite_size_scaling}). This shows that the downstream, memory-free layer exhibits intra-layer scale-free avalanches with an area-controlled cutoff, which is a direct evidence of LRO drag. Furthermore, this downstream scale-invariance emerges even in the presence of quenched disorder, in the form of randomized inter-layer couplings $J_i^{\perp (k)} = \pm J^\perp$, 
and heterogeneous memory timescales in the base layer as shown explicitly in Appendix~\ref{appendix:hetero}.

\subsection{Phase diagram}

\begin{figure*}
    \centering
    \includegraphics[width=1.0\linewidth]{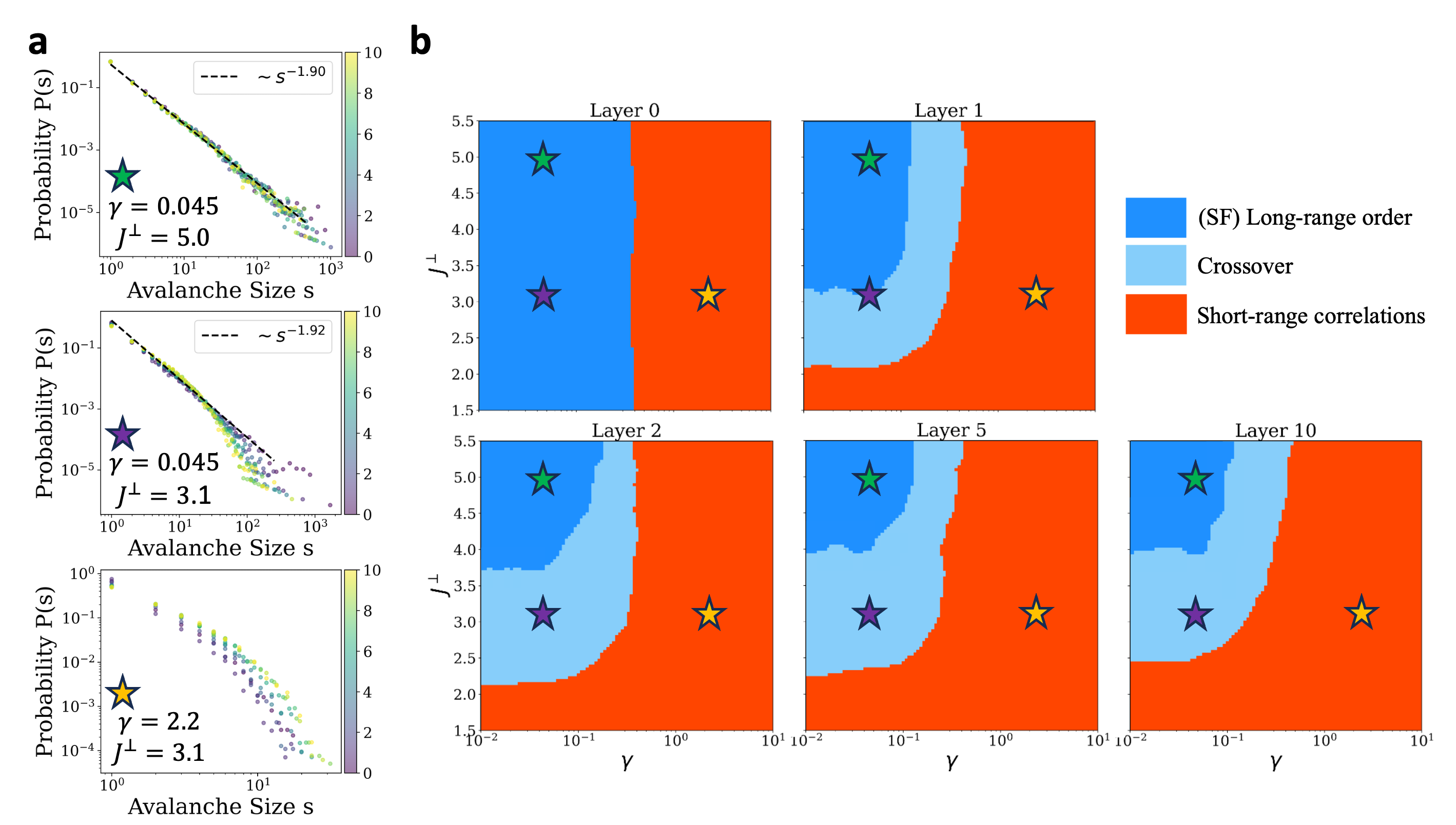}
    \caption{Depth-resolved phase structure and representative avalanche statistics. (a) Avalanche size distributions $P(s)$ for $\{\gamma, J^\perp\} = \{0.045, 5.0\}, \{0.045, 3.1\}, \{2.2, 3.1\}$ (stars correspond to locations in the phase diagram in panel (b). Dashed lines indicate power-law guides with exponents $\gamma\simeq 1.90$ and $1.92$, respectively. (b) Phase diagrams in the $\{\gamma, J^\perp\}$ plane at depths $k=0, 1, 2, 5, 10$. Colors denote long-range order, crossover, and short-range correlations. Stars mark the parameter points used in (a). All points in the parameter space are simulated over 25 instances for $T=200$ simulation time (a.u.) (with initial transient dynamics removed) on a $64 \times 64$ lattice.}
    \label{fig:phase_diagram}
\end{figure*}

Having shown that MILRO can drag to an adjacent layer from the base one, we now want to understand how deep that drag can go, and how it depends on the main parameters in the system, in particular the frequency of the slow (memory) degrees of freedom, $\gamma$, and the strength of the inter-layer coupling, $J_i^{\perp(k)}$. We then consider the base driving layer coupled to ten downstream layers. Scanning the memory frequency scale $\gamma$ and feedforward strength $J^\perp$ yields the depth-resolved phase structure in Fig.~\ref{fig:phase_diagram}: (i) an LRO phase with scale-free avalanches in every layer; (ii) a crossover band where power laws persist but the range shrinks with depth; and (iii) a short-range phase with only small avalanches.

Representative cuts at $\{\gamma, J^\perp\} = \{0.045, 5.0\}$, $\{0.045, 3.1\}$, and $\{2.2, 3.1\}$ illustrate these regimes, with avalanche distributions shown in Fig.~\ref{fig:phase_diagram}(a). When $\gamma$ is sufficiently low (to introduce strongly correlated behavior in the base layer $k=0$) {\it and} $J^\perp > 4$ (to ensure that spins in deeper layers are always highly correlated with analogous spins in the base layer), LRO is detected up to the deepest layers tested. The size of this region appears to saturate, so we anticipate that such LRO behavior should persist in this parameter regime to arbitrarily deep layers. This corroborates our intuition from Sec.~\ref{sec:MILROa}; once $J^\perp/J > 4$, inter-layer interactions will dominate over intra-layer ones {\it at every site}, enabling LRO to be dragged to arbitrary depth without decay (if it exists in the base layer). On the other hand, when $\gamma$ is small but $J^\perp$ is not so large, LRO exists in the base layer, but inter-layer interactions must compete with intra-layer ones. In the regime where $2 < J^\perp/J < 4$, LRO can be dragged between layers, but the effectiveness of this drag phenomenon decays with depth. As more instances of frustration are encountered while moving through to deeper layers, correlations between the $k^\text{th}$ layer and the base layer will weaken with increasing $k$. Finally, if $\gamma$ is so large that LRO is never induced in the base layer at all, then the deeper layers only exhibit short-range correlations, as there is no collective state to ``drag'' at all.

The depth evolution reflects a simple competition. Memory in the base layer generates extended correlations; feedforward couplings transmit them downstream before they decay. In a band $J^\perp/J \in (2,4)$ the inter-layer field is typically stronger than the net in-plane constraint at a site but does not fully dominate.\footnote{As anticipated in Sec.~\ref{sec:MILROa}, the lower bound for $J^\perp/J$ to observe LRO in deeper layers is indeed larger than $3/2$.} In this regime, large-scale avalanches are still generally able to propagate between layers, leading to wide phases of LRO which only shrink slowly when moving between subsequent layers (Fig.~\ref{fig:phase_diagram}).

It is worth mentioning that even the crossover regime, which certainly lacks scale-freeness, can be considered to have an effective long-range interaction, induced by this drag effect (since avalanches with up to a few hundred participating spins exist). This is in stark contrast with the short-ranged region, in which avalanches above size 10 are already quite rare; see Fig.~\ref{fig:phase_diagram}(a). Additionally, this crossover region appears to shrink quite slowly when comparing deeper layers.

Lastly, we must emphasize that for $J^\perp/J > 4$ the inter-layer term {\it always} dominates and downstream layers largely follow the base layer's dynamics, causing inter-layer synchronizations. This is precisely why the LRO region saturates in size after a few layers (see the similarity between Layer 5 and 10 in Fig.~\ref{fig:phase_diagram}(b)). This is also why we expect the phase diagram not to change much if we were to simulate even deeper layers. On the contrary, for $J^\perp/J < 2$ the system cannot overcome typical in-plane basins of attraction and layers quickly freeze. See Appendix~\ref{appendix:spin_flip_time} for a more rigorous analytical understanding of this point, and in Appendix~\ref{appendix:3d_avalanches}, we demonstrate the phase structures when we consider correlations within the entire three-dimensional stack instead of intra-layer only. It is also worth noticing that the transfer of LRO does {\it not} arise trivially from the forcing of spins in lower layers to faithfully track the dynamics of their upper counterparts. We explicitly show this in Appendix~\ref{appendix:no_forcing}.

\subsection{Transient build-up of order}

\begin{figure}
    \centering
    \includegraphics[width=0.9\linewidth]{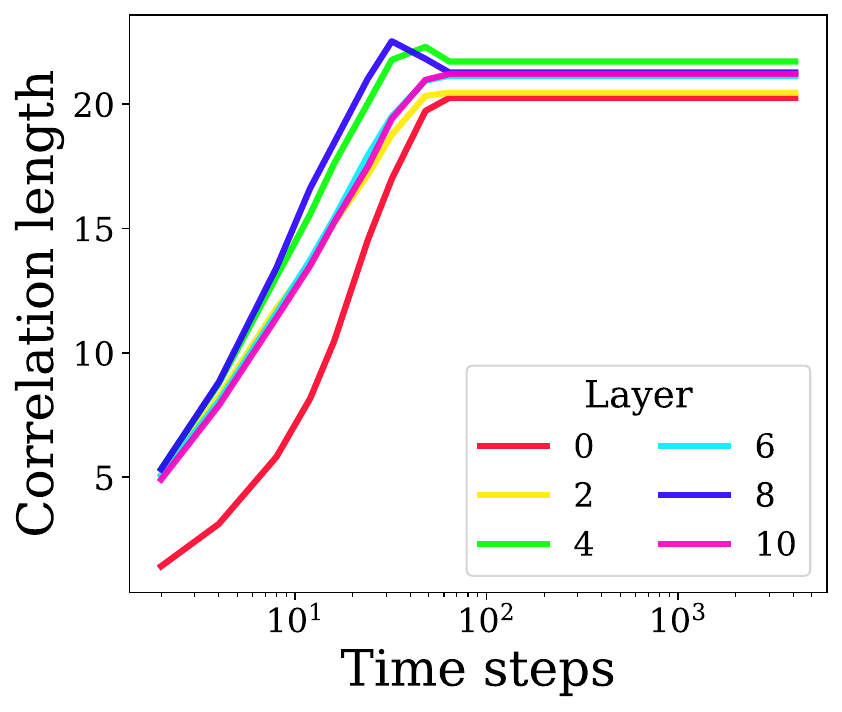}
    \caption{Transient build-up of long-range order across layers. Time evolution of the in-plane correlation length $\xi_k(t)$ for the driving layer ($k=0$) and five downstream layers ($k=2, 4, 6, 8, 10$) at $\{\gamma, J^\perp\} = \{0.045, 5.0\}$ on a $64\times 64$ lattice with $J=1$, simulated over 100 instances for 4096 time steps, where each time step $\Delta t=0.017$. All layers exhibit a rapid rise over $\sim 10^2$ steps followed by a plateau. Mild overshoots and subsequent relaxation reflect the exclusion of system-wide events from $\xi$. Note that this transient timescale is $\sim 1/\gamma$, as expected, since $1/\gamma$ acts as an approximate timescale for the memory dynamics.}
    \label{fig:transient}
\end{figure}

Finally, to visualize how drag develops in time across layers, we first initialize each layer in its ground state and track the correlation length $\xi_k(t)$, defined as the mean in-plane separation of two events belonging to the same avalanche (excluding system-spanning avalanches to avoid trivial divergences, in analogy with percolation observables~\cite{christensen2002percolation}) in layer $k$, from the start of the simulation up to time $t$. As shown in Fig.~\ref{fig:transient}, for $\{\gamma, J^\perp\} = \{0.045, 5.0\}$ on a $64\times 64$ lattice, $\xi_k(t)$ rises rapidly over $\sim 10^2$ time steps and then plateaus in all layers. For different values of $k$, all $\xi_k(t)$ are close to each other at given $t$, consistent with the inter-layer synchronization regime discussed above. Mild oscillations of $\xi(t)$ in deeper layers arise when growing avalanches begin to hit system size and are excluded from the estimator, producing a slight decrease toward the final plateau. 

\section{Conclusion}

We have shown that slow memory degrees of freedom can seed long-range order (LRO) in a locally coupled layer and, via purely local feedforward couplings, drag this order into downstream layers that possess neither memory nor long-range interactions. Across a broad region of parameters, the driven layers display power-law avalanche statistics with finite-size scaling indistinguishable from the driving layer, establishing MILRO drag as a robust, interaction-mediated propagation of order rather than a transport effect. 

These results suggest practical principles for multilayer neuromorphic systems: a memory-rich ``source'' layer can broadcast coherent, scale-free dynamics to passive layers without fine tuning. This mechanism further provides a concrete hypothesis for inter-laminar correlations in the brain cortex, readily testable with existing volumetric neural imaging techniques \cite{rabut20194d, demas2021high, gutierrez2022unique}: slow adaptive processes, confined to specific layers, can sustain depth-wide coherence through local pathways.  Based on these results, we expect this MILRO drag effect could be not just a mechanism of theoretical interest but also a design tool for layered, adaptive matter and computation.

\section*{Acknowledgments}
We thank Kaushik Madapati for useful discussions. This work was funded by the National Science Foundation via grant No. ECCS-2229880. M.D. also acknowledges funding by the Alexander von Humboldt Stiftung through the Humboldt Research Award.

\section*{Author contributions}
M.D. suggested and supervised the work. Y.-H.Z. and C.S. contributed equally to this work. Y.-H.Z. performed many numerical simulations, and C.S. the theoretical analysis and part of the numerical simulations. All authors have read and contributed to the writing of the paper.

\appendix

\section{Numerical details}
\label{appendix:numerics}

We simulate Eqn.~\ref{eq:model} on $k$ square layers of size $L\times L$ stacked along the feedforward direction. Unless stated otherwise, open boundaries are used in-plane and all parameters are homogeneous within a layer. Time evolution is integrated with a fixed–step, fourth–order Runge–Kutta scheme using $\Delta t = 0.048\,\gamma^{1/3}$, which resolves both the fast spin dynamics and the slow memory dynamics over the range of $\gamma$ explored. To eliminate transients, we discard the first $10^3$ steps and record the subsequent trajectory for avalanche analysis. For each $\{\gamma,J^\perp\}$, avalanches are collected over multiple disorder realizations of $\{J_{ij}^{(k)},J_i^{\perp(k)}\}$ and independent random seeds. 

An \emph{avalanche} is a connected burst of spin flips in space and time. We detect flips by monitoring sign changes of $s_i^{(k)}(t)$; each flip is stored with its lattice site and timestamp. Two flips are deemed connected if they occur on nearest–neighbor sites and within a temporal proximity window $\Delta t_w$. We construct avalanches by recursively merging connected flips (single–linkage clustering on the event graph). Avalanche size $s$ is the number of flips in the cluster. All distributions in Figs.~\ref{fig:finite_size_scaling} and \ref{fig:phase_diagram} are obtained with this event–linked procedure.

The event–linked algorithm scales poorly at large $L$ or when activity is dense, because the number of potential space–time links grows rapidly. For heavier workloads we therefore use a coarse–grained alternative: we discretize time into bins of width $\Delta t_w$ and mark a site as ``active'' in bin $n$ if at least one flip occurs at that site in $[n\Delta t_w,(n{+}1)\Delta t_w)$. This maps detection to a 3D (2D space $+$ time) site–percolation problem with lattice spacing $(1,1,\Delta t_w)$ and diagonal connections; avalanches correspond to connected clusters in this 3D grid.

\begin{figure}[t]
    \centering
    \includegraphics[width=0.98\columnwidth]{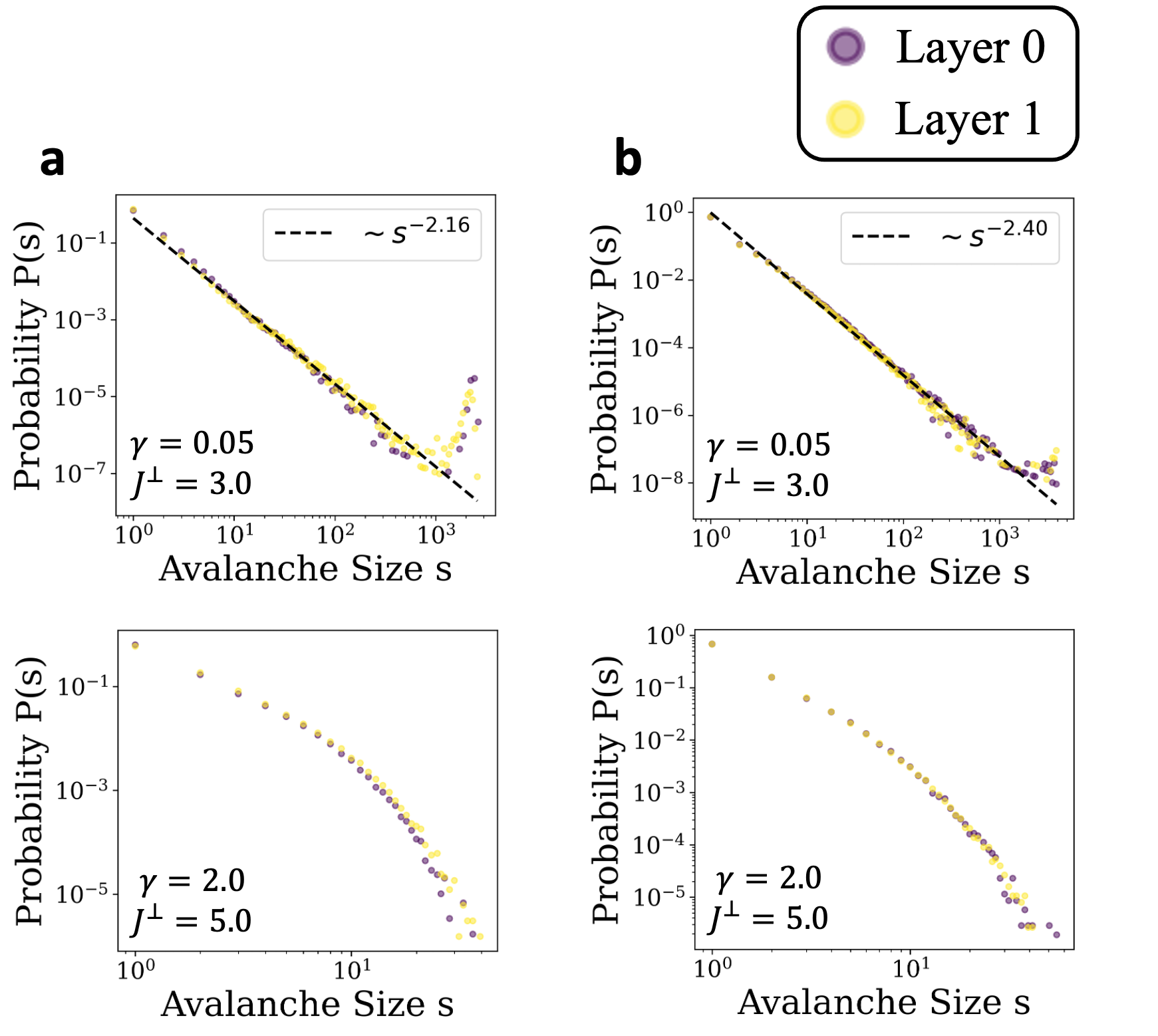}
    \caption{Comparison of avalanche distributions extracted using the event-linked algorithm (a) and the coarse-grained alternative (b). Both extraction procedures produce similar phase structure, including both following power-law distributions in the LRO regime (e.g., $\{\gamma, J^\perp\} = \{0.05, 3.0\}$) and both only yielding small-scale avalanches in the short-range correlated regime (e.g., $\{\gamma, J^\perp\} = \{2.0, 5.0\}$). For each distribution, $1000$ instances are simulated over $T=200$ simulation time (a.u.) on a $64 \times 64$ lattice, with $\Delta_{tw}/dt = 75$.}
    \label{fig:extraction_comparison}
\end{figure}

Across all parameter sets tested, the two methods yield almost identical power–law regimes and consistent cutoffs within statistical error. However, explicit time binning breaks continuous time translation invariance and can introduce slightly different power-law exponents, as well as artifacts near the tail. We therefore report all main text results using the more physically reasonable event–linked algorithm and reserve the binned 3D percolation method for large sweeps where its lower complexity is advantageous. See Fig.~\ref{fig:extraction_comparison} for a comparison of these two avalanche extraction approaches.

\section{Choosing the temporal proximity window}\label{appendix:window}

The avalanche definition involves a single free parameter, the temporal proximity window $\Delta t_w$. As with other common avalanche definitions (e.g., thresholding or fixed-time binning)~\cite{beggs2003neuronal,bearden2019critical}, results are typically robust over a finite range of this parameter. Here we describe our procedure for selecting $\Delta t_w$ in a data-driven and physically motivated manner.

Because memory evolves on the slow timescale $1/\gamma$ and inter-layer transmission is controlled by $J^\perp$, the optimal $\Delta t_w$ should be large enough to combine causally related flips within a burst, yet small enough to avoid merging distinct events. We therefore choose $\Delta t_w$ to \emph{maximize} an in-plane correlation length $\xi(\Delta t_w)$, defined as the mean spatial separation of two events within the same avalanche (excluding system-spanning avalanches to prevent a trivial divergence). This criterion favors windows that preserve the largest coherent structures without over-aggregation.

\begin{figure*}[t]
    \centering
    \includegraphics[width=\linewidth]{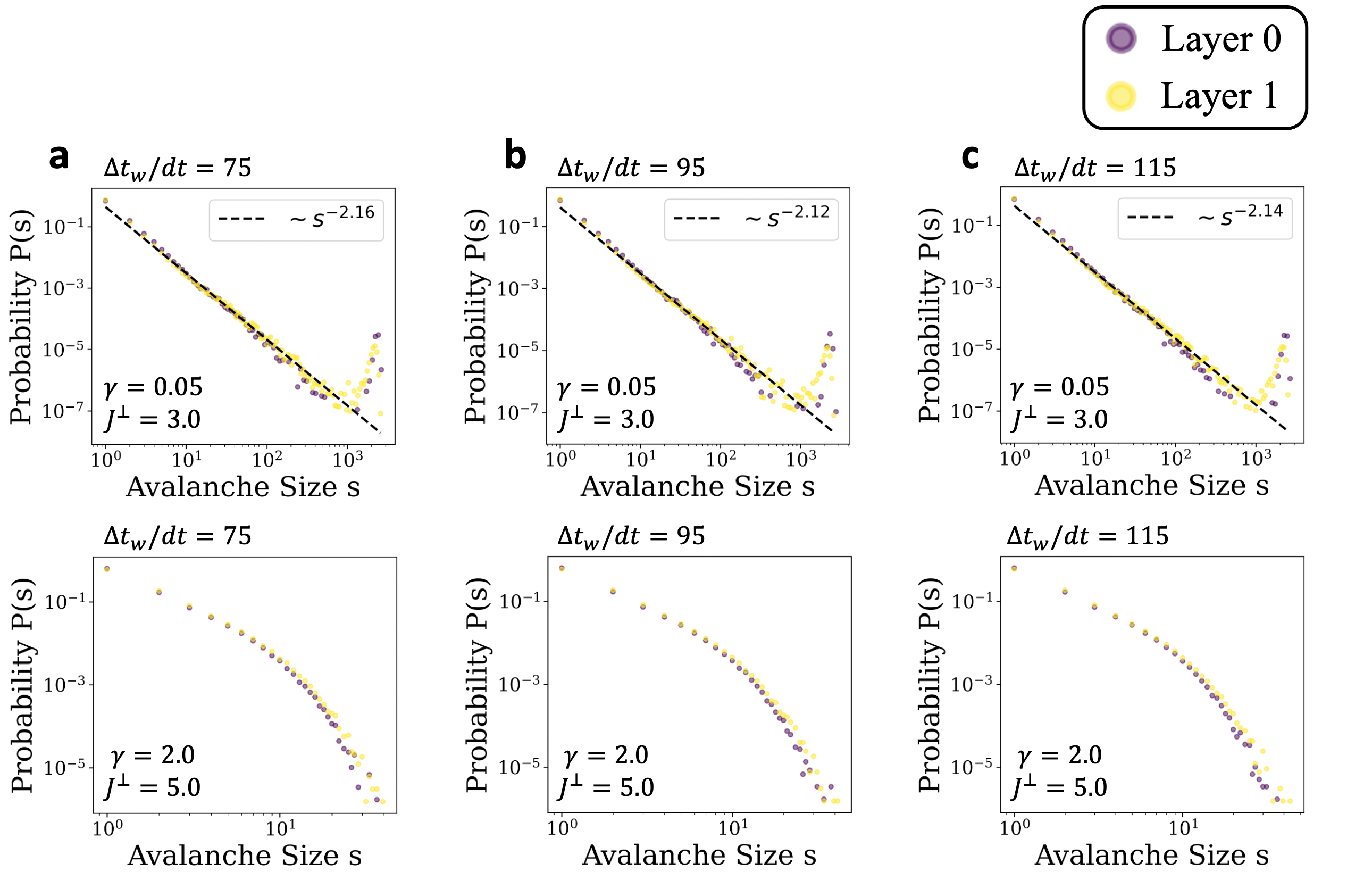}
    \caption{Avalanche distributions over a variety of time windows. Avalanches are extracted at both $\{\gamma, J^{\perp}\} = \{0.05, 3.0\}$ and $\{2.0, 5.0\}$ with a maximum of $75$, $95$, and $115$ integration timesteps (in subplots a, b, and c, respectively) between successive flips in a single avalanche (the number of integration timesteps can be scaled by the simulation timestep $dt$ to find $\Delta_{tw}$ in physical, simulation time). This region represents a plateau in $\Delta_{tw}$-space at which different time windows yield nearly identical distributions at both points in physical parameter space, suggesting the time window does not need to be fine-tuned to yield the distributions we observe. Avalanches are extracted over 1000 instances, each simulated for $T=200$ simulation time (a.u.), on a $64 \times 64$ lattice.}
    \label{fig:tw_example}
\end{figure*}

\begin{figure*}[htbp]
    \centering
    \includegraphics[width=\linewidth]{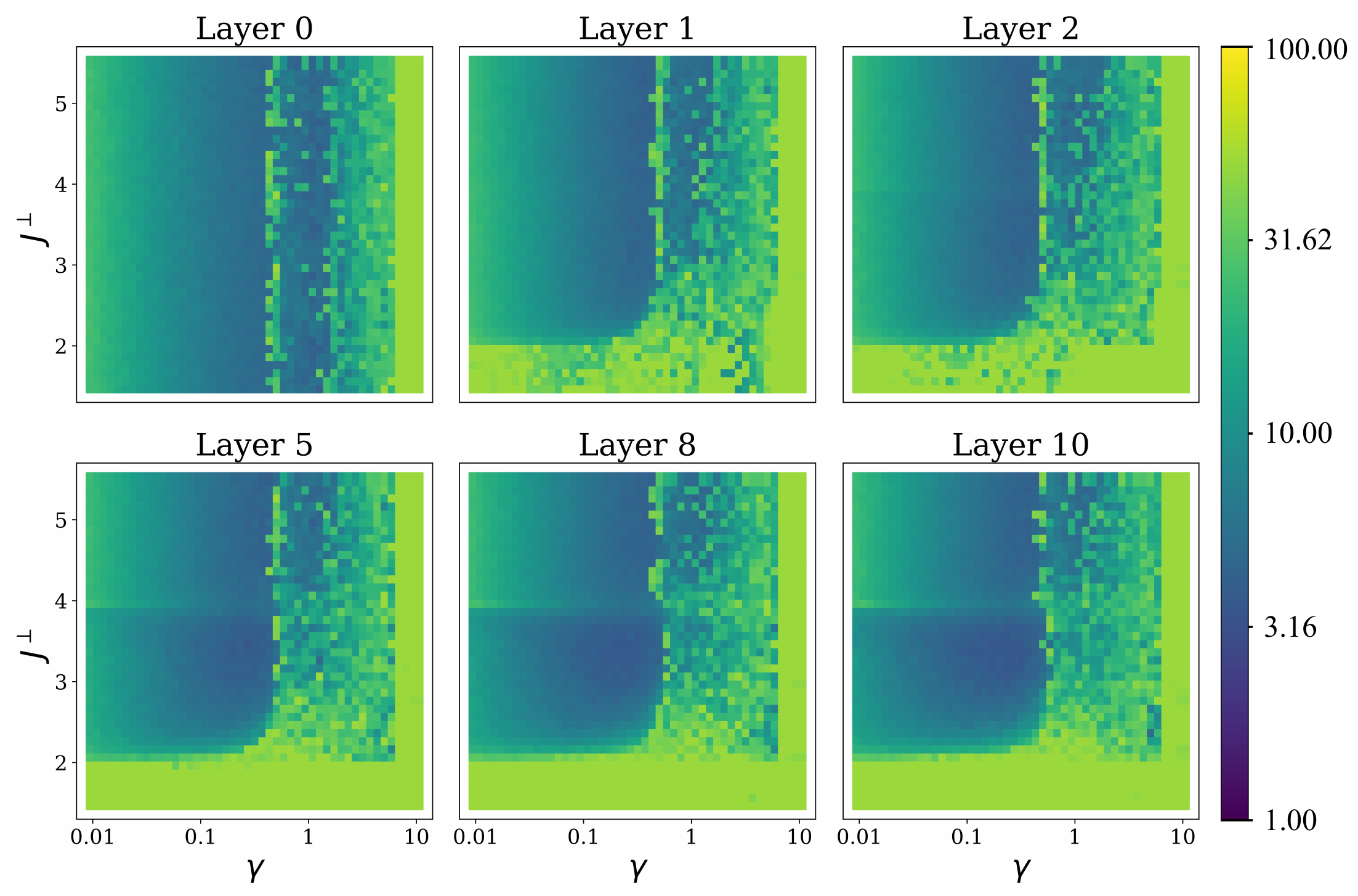}
    \caption{The optimal temporal proximity windows $\Delta t_w$ for different parameter configurations. Each data point is obtained by simulating the dynamics of 1 instance on a $64\times 64$ lattice over 8000 time steps ($T=8000\Delta t$, $\Delta t=0.048 \gamma^{1/3}$), extracting the avalanches using 20 different values of $\Delta t_w$ and selecting the one that maximizes the correlation length. We can loosely see the phase structures presented in Fig.~\ref{fig:phase_diagram}. We also note a line of large $\Delta t_w$ at $J^\perp=4$, corresponding to the diverging average spin flip time that will be derived in Appendix~\ref{appendix:spin_flip_time}.}
    \label{fig:optimal_tw}
\end{figure*}

Figure~\ref{fig:tw_example} shows $P(s)$ for several $\Delta t_w$ at $\{\gamma, J^\perp\} = \{0.05, 3.0\}$ (a point in the LRO regime) and $\{2.0, 5.0\}$ (a point in the short-range correlated regime) on a $64\times 64$ lattice. Only very minor changes are visible in the distributions as $\Delta t_w$ is varied, suggesting that a broad plateau of near-optimal windows yields indistinguishable power-law regimes and essentially identical exponents. We observe similar plateaus at many other points in parameter space. We use the maximizer of $\xi(\Delta t_w)$ within these plateaus to select an ``optimal'' window $\Delta_{tw}$.

This optimal $\Delta t_w$ depends on $\{\gamma,J^\perp\}$: faster memory ($\gamma$ large) typically favors smaller windows, whereas stronger drag ($J^\perp$ large) can tolerate slightly larger windows due to tighter cross-layer coupling. For each parameter pair we sweep $\Delta t_w$ over a wide range, identify the maximizer of $\xi(\Delta t_w)$, and verify that the inferred exponent $\tau$ is stable across the plateau.

The resulting optimal windows are summarized in Fig.~\ref{fig:optimal_tw}. All phase-diagram measurements in Fig.~\ref{fig:finite_size_scaling} and Fig.~\ref{fig:phase_diagram} use these $\Delta t_w$ values.

\section{Heterogeneous Memory Timescales}\label{appendix:hetero}

To test how the emergence of LRO drag is influenced by the presence of spatially heterogeneity in the base layer, we perform an additional set of simulations in which the memory parameter $\gamma$ in Eqn.~\ref{eq:model} now takes a unique value, $\gamma_{ij}$, for each edge in the base layer, rather than this constant being fixed at each edge. These parameters $\gamma_{ij}$ are independently sampled from a Gaussian distribution with mean $\mu_\gamma$ and standard deviation $\sigma_\gamma$. Parameters $\gamma_{ij}$ are subsequently bounded from below by $10^{-3}$ to assure that these frequencies remain positive. Note that each memory variable still evolves according to a dynamical equation with the same functional form given in Eqn.~\ref{eq:model}, but $\gamma$ now takes a distinct value at each edge.

\begin{figure}[htbp]
    \centering
    \includegraphics[width=0.98\columnwidth]{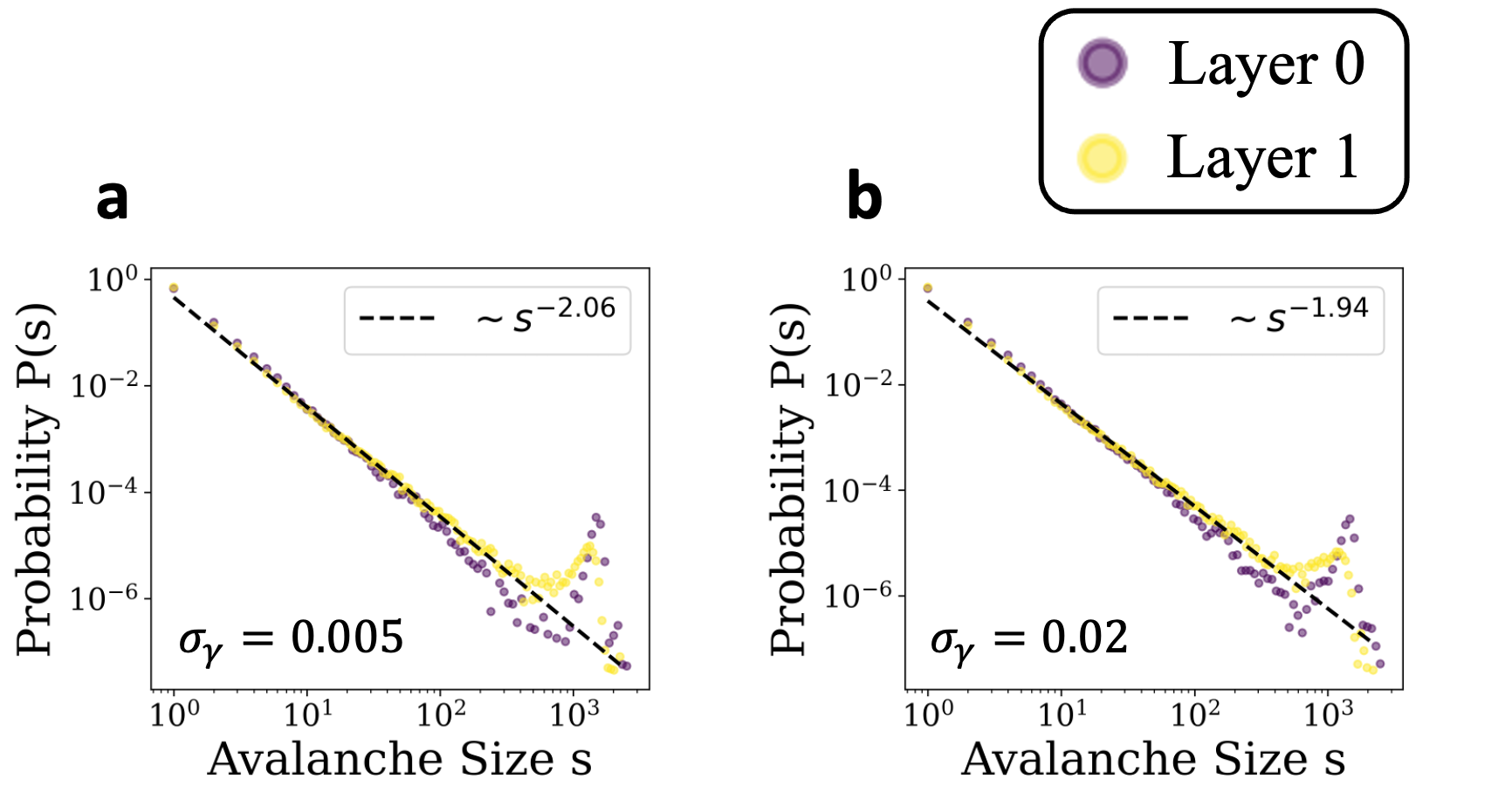}
    \caption{Avalanche distributions with randomized memory parameter $\gamma_{ij}$ in the base layer, where each $\gamma_{ij}$ is independently drawn from a Gaussian with mean $\mu_\gamma = 0.05$ and standard deviation $\sigma_\gamma$. Avalanches are extracted at both $\sigma_\gamma = 0.005$ (a) and $\sigma_\gamma = 0.02$ (b). In each case, scale-free avalanches persist in both the base layer 0 and layer 1, despite the heterogeneity of the memory timescales. Avalanches are extracted over 1000 instances, each simulated for $T=200$ simulation time (a.u.), on a $64 \times 64$ lattice with $\Delta_{tw}/dt = 125$, and $J^{\perp}=3.0$.}
    \label{fig:hetero_memories}
\end{figure}

Fig.~\ref{fig:hetero_memories} shows the resulting avalanche distributions in layers 0 and 1 for Gaussians with mean $\mu_\gamma = 0.05$ and two different standard deviations $\sigma_\gamma = 0.005$ and $\sigma_\gamma = 0.02$. In both cases, scale-free avalanche distributions in layer 0 are transferred to layer 1 despite the fact that individual memories evolve over different dynamical timescales. This further suggests that no fine-tuning is present in our model, and that no precise functional form of the memories is required to generate or transfer LRO. So long as the majority of memories evolve much more slowly than their accompanying spins (which is guaranteed so long as $\mu_\gamma$ is not too large), LRO will be produced that can then be dragged to subsequent layers.

\section{Analytical Derivation of Average Spin Flipping Time}
\label{appendix:spin_flip_time}

To clarify the relevant timescales involved in our model, we will explicitly calculate the time $\Delta t$ over which a single spin $s_i^{(k)}$ flips, in both the base layer $k = 0$ and deeper layers $k \geq 1$. We will assume that all neighbors of the flipping spin, $s_j^{(k)}$ and $s_i^{(k - 1)}$, are constant over this time interval, and that $s_i^{(k)}$ changes approximately linearly over this flipping interval. Furthermore, without loss of generality, we will assume that $s_i^{(0)}$ flips from $-1$ to $+1$, so that $\Delta s_i^{(0)} = 2$. Note that the analysis would proceed identically, up to some factors of $-1$, if this flip was from $+1$ to $-1$.

First, since the change in a spin's magnitude must be $2$, we can relate the flipping time $\Delta t$ to the derivative $\dot{s}_i^{(k)}$:

\begin{equation}
    s_i^{(k)}(\Delta t) = s_i^{(k)}(0) + \dot{s}_i^{(k)} \cdot \Delta t \implies \dot{s}_i^{(k)} \cdot \Delta t = 2.\label{eq:flipping_time}
\end{equation}

From here, we can use the functional form for $\dot{s}_i^{(k)}$, in Eqn.~\ref{eq:model}, to estimate $\Delta t$. We reiterate it here:

\begin{equation}
\begin{aligned}
    \dot{s}_i^{(0)}=& \sum_{\langle ij \rangle} J_{ij}^{(0)}s_j^{(0)}-g\sum_{\langle ij\rangle}x_{ij}s_i^{(0)},\\
    \dot{s}_i^{(k)}=& \sum_{\langle ij \rangle} J_{ij}^{(k)}s_j^{(k)} + J_i^{\perp(k)}s_{i}^{(k-1)} \qquad (k\ge 1),\\
    \dot{x}_{ij} =& \gamma(C_{ij}-\delta), \quad C_{ij}=(J_{ij}^{(0)}s_i^{(0)} s_j^{(0)} + 1) / 2.
\end{aligned}
\end{equation}

Let us consider the base layer case, $k = 0$, first. There, the two types of interactions are the intra-layer spin-glass interactions and the memory interactions. Since $x_{ij}$ and $s_i^{(0)}$ are both dynamical, we can estimate their contributions to $\dot{s}_i^{(k)}$ by taking the average value of their product $\overline{x_{ij}s_i^{(0)}}$:

\begin{equation}
    \dot{s}_i^{(0)}\approx \sum_{\langle ij \rangle} J_{ij}^{(0)}s_j^{(0)}-g\sum_{\langle ij\rangle}\overline{x_{ij}s_i^{(0)}}.
\end{equation}

Since $s_i^{(k)}$ varies linearly in time (and we have assumed all nearest-neighbors $s_j$ to be fixed during this interval of width $\Delta t$), $x_{ij}(t)$ will follow some parabolic trajectory, which can be solved for:

\begin{equation}
\begin{aligned}
    \dot{x}_{ij} =& \frac{\gamma}{2}J_{ij}^{(0)} \bigg( 
    \frac{2}{\Delta t}t - 1\bigg) s_j^{(0)} - \gamma (\delta - 1/2),\\
    \implies x_{ij} =& \int dt \, \Bigg( \frac{\gamma J_{ij}^{(0)}s_j^{(0)}}{\Delta t}t + \gamma \bigg( -\frac{J_{ij}^{(0)} s_j^{(0)}}{2} - \delta^* \bigg) \Bigg),\\
    =& \frac{\gamma J_{ij}^{(0)} s_j^{(0)}}{2 \Delta t} t^2 + \Bigg( -\frac{\gamma J_{ij}^{(0)} s_j^{(0)}}{2} - \gamma \delta^* \Bigg)t + x_{ij}(0),\\
    \equiv& \alpha t^2 + \beta t + x_{ij}(0).
\end{aligned}
\end{equation}

\noindent where $\delta^* \equiv \delta - 1/2$. Now that we have an expression for $x_{ij}$(t), we can perform a simple integration to find $\overline{x_{ij}s_i^{(0)}}$:

\begin{equation}
\begin{aligned}
    \overline{x_{ij}s_i^{(0)}} =& \frac{1}{\Delta t} \int_0^{\Delta t} dt \, x_{ij}s_i^{(0)},\\
    =& \frac{1}{\Delta t} \int_0^{\Delta t} dt \, \bigg( \alpha t^2 + \beta t + x_{ij}(0) \bigg)\bigg( \frac{2}{\Delta t}t - 1 \bigg),\\
    =& \frac{1}{\Delta t} \Bigg( \frac{2 \alpha}{4 \Delta t}(\Delta t)^4 + \bigg( \frac{2 \beta}{3 \Delta t} - \frac{\alpha}{3} \bigg)(\Delta t)^3 + \bigg( \frac{2 x_{ij}(0)}{2 \Delta t} - \frac{\beta}{2} \bigg)(\Delta t)^2 - x_{ij}(0)(\Delta t) \Bigg),\\
    =& \frac{\alpha}{6}(\Delta t)^2 + \frac{\beta}{6}(\Delta t),\\
    =& \frac{\Delta t}{6} \bigg( \frac{\gamma J_{ij}^{(0)}s_j^{(0)}}{2} - \frac{\gamma J_{ij}^{(0)}s_j^{(0)}}{2} - \gamma \delta^* \bigg) = - \frac{\gamma \delta^* \Delta t}{6},
\end{aligned}
\end{equation}

\noindent giving us a more concrete expression for $\dot{s}_i^{(0)}$ which now only depends on free parameters and $\Delta t$.

\begin{figure*}[t]
    \centering
    \includegraphics[width=\linewidth]{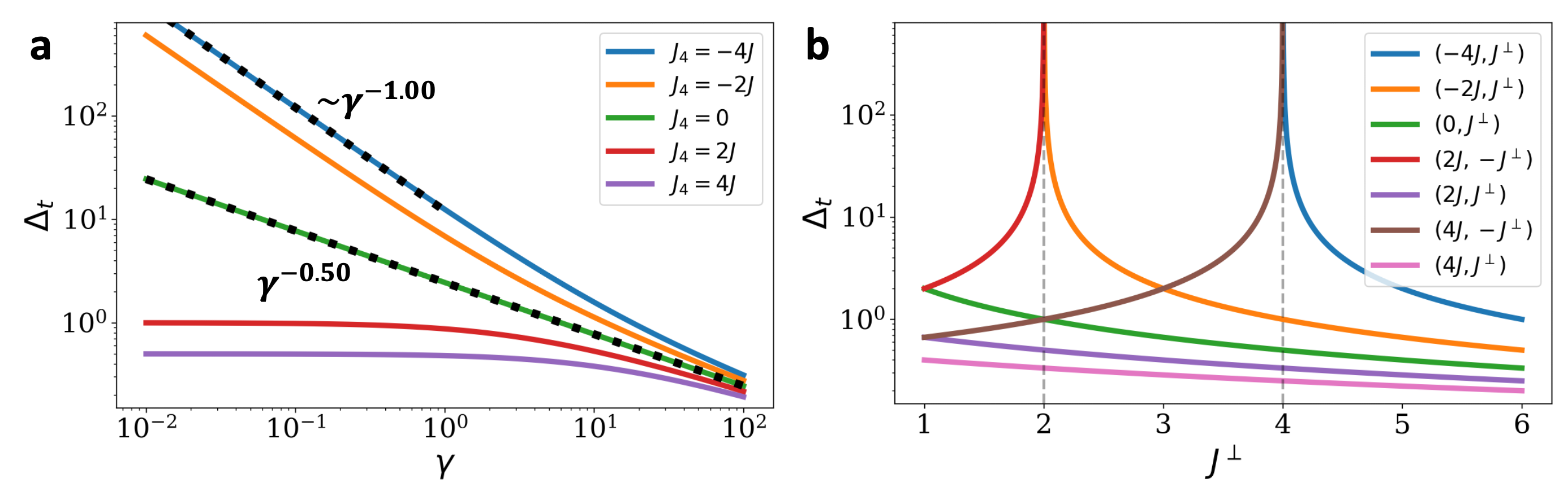}
    \caption{The flipping time $\Delta t$ over which a single spin $s_i^{(k)}$ flips, in both (a) the base layer $k = 0$ and (b) other layers $k \geq 1$. This timescale varies depending on the nature of the relevant spin-glass interactions. (a) When $J_4 = 0$, $\Delta t = \sqrt{3/g \gamma \delta^*}$. When $J_4 < 0$, the flipping time decays as $\Delta t \sim 1/\gamma$, until $\gamma \sim O(1)$. When $J_4 > 0$, the flipping time remains constant at values near $1$ for $J_4 = 2J$ and $1/2$ for $J_4 = 4J$, until $\gamma \sim O(1)$. (b) $\Delta t$ decays monotonically when both $J_4 > 0$ and $J_1^\perp > 0$. Divergences appear at $J^\perp = 2J$ and $J^\perp = 4J$ when the inter-layer and intra-layer spin-glass interactions are in opposition, and in these cases, flips can only occur within a particular range of $J^\perp$. When $J_4 < 0$ and $J_1^\perp < 0$, no flip will ever occur, so these curves cannot be plotted. Parameters chosen: $g = 2$, $\delta^* \equiv (\delta - 1/2) =  1/4$, $J = 1$.}
    \label{fig:SM_fig_1}
\end{figure*}

We can now write Eqn.~\ref{eq:flipping_time} as a quadratic equation for $\Delta t$:

\begin{equation}
\begin{aligned}
    \frac{2}{\Delta t} =& \sum_{\langle ij \rangle} \bigg[ J_{ij}^{(0)}s_j^{(0)} + \frac{g \gamma  \delta^* \Delta t}{6} \bigg],\\
    \implies 0 =& \frac{2 g \gamma \delta^*}{3} (\Delta t)^2 + J_4 (\Delta t) - 2,\\
    \implies \Delta t =& \frac{3}{4 g \gamma \delta^*} \bigg(-J_4 + \sqrt{J_4^2 +\frac{16 g \gamma \delta^*}{3}}\bigg),\label{eqn:k_0_delta_t}
\end{aligned}
\end{equation}

\noindent where $J_4 \equiv \sum_{\langle ij \rangle} J_{ij}^{(k)} s_j^{(k)}$. As in the main text, the four nearest-neighbor $J_{ij}^{(0)} s_j^{(0)}$'s will each take values $\{ -J, J \}$. Thus, $J_4 \equiv \sum_{\langle ij \rangle} J_{ij}^{(0)} s_j^{(0)}$ will take the values $\{-4J, -2J, 0, 2J, 4J\}$. In Fig.~\ref{fig:SM_fig_1}(a) we plot this flipping time $\Delta t$ for each of these five cases as a function of $\gamma$, with $g=2$, $\delta=3/4$, and $J=1$ fixed. As expected, $\Delta t$ always decreases monotonically as a function of $\gamma$, since the faster the memory frequency $\gamma$, the faster $s_i^{(k)}$ will be able to flip. All cases tend towards the power law decay $\Delta t \sim \gamma^{-1/2}$ in the large-$\gamma$ limit, when the memory interactions will dominate and the strength of the spin-glass interactions becomes irrelevant. Cases where $J_4 < 0$ follow $\Delta t \sim 1/\gamma$ for smaller $\gamma$. This makes sense, as these configurations would be stable in the absence of memories, so the slowly-varying memories must oppose the spin-glass interactions for a longer duration before a flip can be completed. On the other hand, $J_4 > 0$ cases tend towards constant $\Delta t$ at small $\gamma$, as these flips are almost entirely instigated by the spin-glass interactions themselves if $\gamma$ is not too large. The boundary case, $J_4 = 0$, corresponds to a flip in $s_i^{(0)}$ leaving the energy of the spin-glass configuration (ignoring the memory variables) unchanged. In this case, the solution to Eqn.~\ref{eqn:k_0_delta_t} takes the simple form

\begin{equation}
    \Delta t = \sqrt{\frac{3}{g \gamma \delta^*}}.
\end{equation}

\begin{figure*}[htbp]
    \centering
    \includegraphics[width=\linewidth]{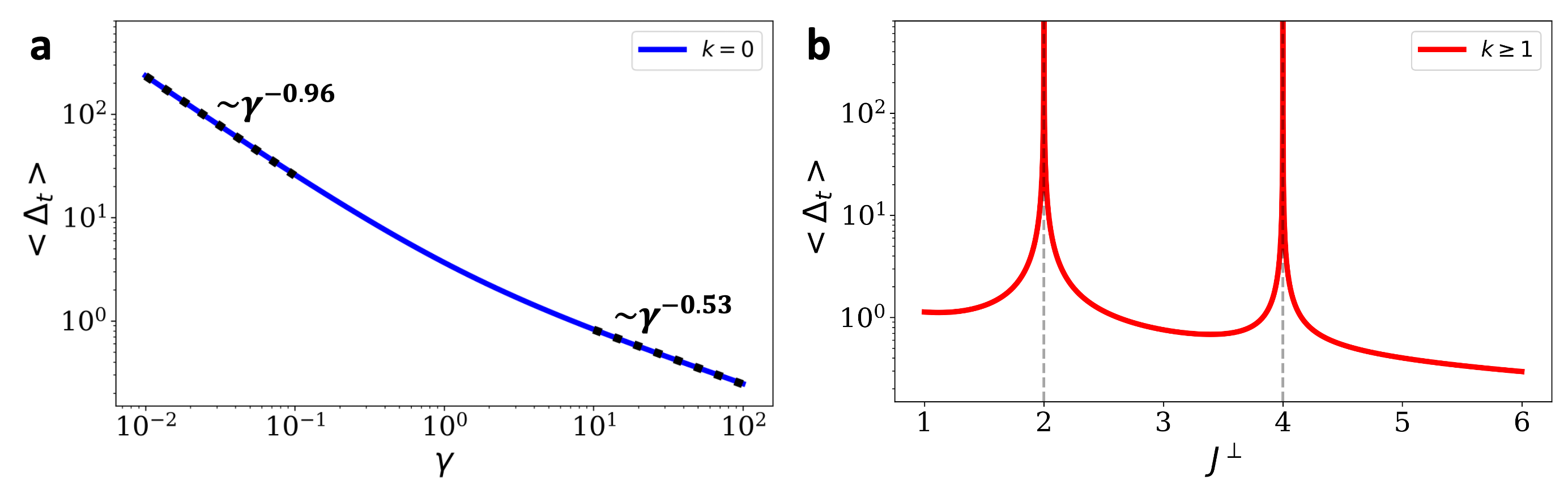}
    \caption{The average time $\Delta t$ over which a single spin $s_i^{(k)}$ flips, in both (a) the base layer $k = 0$ and (b) other layers $k \geq 1$. In both cases, all $\langle\Delta t\rangle$ shown in Fig.~\ref{fig:SM_fig_1} are averaged according to the relative likelihood of occurrence, when both $J_{ij}^{(0)}s_i^{(0)}$ and $J_i^{\perp(k)}s_i^{(k - 1)}$ are taken to be drawn from uniform distributions $\{-J, J\}$ and $\{ -J^\perp, J^\perp \}$, respectively. (a) The curve $\langle \Delta t \rangle$ is fitted to a polynomial decay for $\gamma \in [10^{-2}, 10^{-1}]$, yielding $\Delta t \sim \gamma^{-0.96}$, and $\gamma \in [10^1, 10^2]$, yielding $\Delta t \sim \gamma^{-0.53}$. If this plot were to be extended, $\langle \Delta t \rangle$ would tend further towards $\Delta t \sim \gamma^{-1}$ for lower $\gamma$ and $\Delta t \sim \gamma^{-1/2}$ for higher $\gamma$. Parameters chosen: $g = 2$, $\delta^* \equiv (\delta - 1/2) =  1/4$, $J = 1$.}
    \label{fig:SM_fig_2}
\end{figure*}

We can take this one step further by averaging over the entire lattice to find an average flipping time $\langle \Delta t \rangle$, considering that $J_4$ is itself a random variable on the lattice. As in the main text, we will draw the four nearest-neighbor $J_{ij}^{(0)} s_j^{(0)}$'s from the {\it uniform} distribution $\{-J, J\}$ so that $J_4$ takes the values $\{-4J, -2J, 0, 2J, 4J\}$ with probabilities $\{1/16, 4/16, 6/16, 4/16, 1/16\}$. This enables us to take the average value of any function of $J_4$ which appears in the numerical expression for $\Delta t$ to find the average flipping time $\langle \Delta t \rangle$. See Fig.~\ref{fig:SM_fig_2}(a) for a plot of this average flipping time. Unsurprisingly, it is fitted well by a polynomial decay which begins near $\langle \Delta t \rangle \sim \gamma^{-1}$ for small $\gamma$ and then gradually becomes more shallow until $\langle \Delta t \rangle \sim \gamma^{-1/2}$ for sufficiently large $\gamma$.

This procedure can also be applied to other layers with $k \geq 1$. A similar expression to Eqn.~\ref{eq:flipping_time} holds, but this time $\dot{s}_i^{(k)}$ takes a different functional form:

\begin{equation}
    \dot{s}_i^{(k)} = \sum_{\langle ij \rangle} J_{ij}^{(k)}s_j^{(k)} + J_i^{\perp(k)}s_i^{(k - 1)}.
\end{equation}

Again, we can now write an equation for $\Delta t$:

\begin{equation}
    \frac{2}{\Delta t} = \sum_{\langle ij \rangle} J_{ij}^{(k)}s_j^{(k)} + J_i^{\perp(k)}s_i^{(k - 1)} \implies \Delta t = \frac{2}{J_4 + J^\perp_1},\label{eqn:k_1+_delta_t}
\end{equation}

\noindent where $J_4$ is defined as above and $J^\perp_1 \equiv J_i^{\perp (k)} s_i^{(k - 1)} = \{-J^\perp, J^{\perp}\}$, with probabilities $\{1/2, 1/2\}$. We plot these flipping times as a function of $J^\perp$ for each spin-glass configuration in Fig.~\ref{fig:SM_fig_1}(b). We notice quickly that the behavior of $\Delta t(J^\perp)$ depends strongly on the relevant spin-glass configuration. Cases $\{J_4, J_1^\perp\} = \{-4J, -J^\perp\}$, $\{-2J, -J^\perp\}$, or $\{0, -J^\perp\}$ are not plotted, as no flips would occur in any of these cases. When $J_4$ and $J_1^\perp$ are instead strictly {\it positive}, $\Delta t$ decays monotonically, and in these cases, its value is dramatically lower than in the base layer $k = 0$ for most values of $\gamma$. However, when one of $J_4$ and $J_1^\perp$ is positive and one is negative, flipping only occurs in specific regimes. Divergences appear at $J^\perp = 2J$ and $J^\perp = 4J$, corresponding to transition points beyond which either $J_4$ or $J_1^\perp$ become too weak to initiate a flip. Interestingly, $J^\perp = 2J$ and $J^\perp = 4J$ both seem to correspond well with certain phase boundaries in Fig.~\ref{fig:phase_diagram}. When $J^\perp < 2 J$, the inter-layer interaction is consistently weaker than the intra-layer one, yielding short-range effective interactions in layers $k \geq 1$, regardless of the memory effect in layer $k=0$. When $J^\perp > 4 J$, a flip in an adjacent layer will always induce a flip relatively quickly, enabling the collective state to be dragged to arbitrarily an deep layer.\footnote{As mentioned in Sec.~\ref{sec:numerics}, the size of the LRO region in phase space seems to saturate, persisting in the deepest layers simulated when $\gamma < 0.5$ and $J^\perp > 4$.}

Lastly, we can take the average value of the expression in Eqn.~\ref{eqn:k_1+_delta_t}, given all possible values of $J_4$ and $J_1^{\perp}$, to produce an expression for $\langle \Delta t \rangle$. This average flipping time $\langle \Delta t \rangle$ as a function of $J^\perp$ is plotted in Fig.~\ref{fig:SM_fig_2}(b), with $J=1$ fixed. As expected, $\langle \Delta t \rangle$ is less than $1$ when $J^\perp$ is not too close to $2 J$ or $4 J$, but near these values, the flipping timescale diverges.

\section{Three–dimensional avalanche structures}
\label{appendix:3d_avalanches}

The analysis in the main text focused on \emph{in–plane} avalanches, i.e., clusters built from nearest–neighbor connectivity within a single layer. To probe {\it inter}-layer correlations directly, we extend the definition of an avalanche to \emph{three dimensions}: two flip events are considered connected if they occur on nearest–neighbor sites in the full $(x,y,k)$ lattice and within the temporal window $\Delta t_w$ (cf. Appendix~\ref{appendix:numerics}). Therefore clusters may percolate both laterally and across layers.

\begin{figure*}[htbp]
    \centering
    \includegraphics[width=\linewidth]{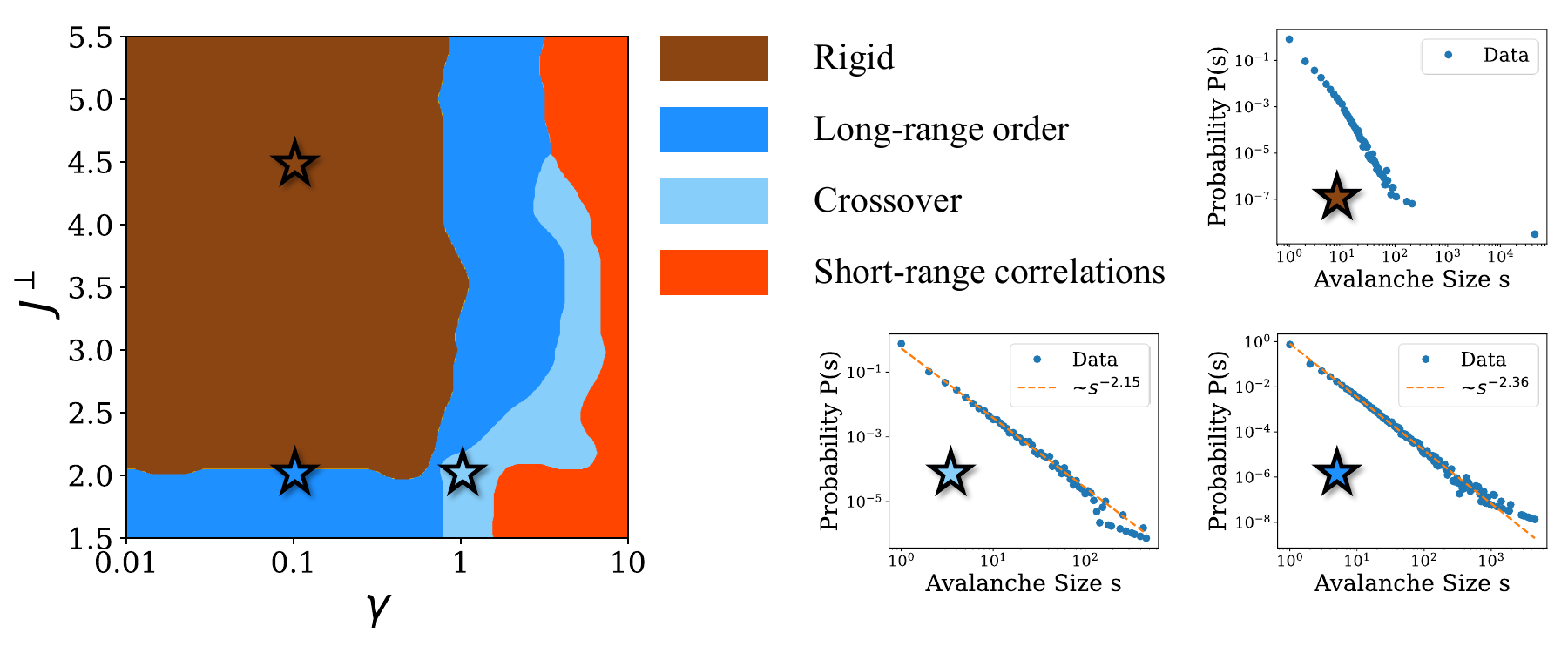}
    \caption{Three-dimensional avalanche structure and phase map. Left: Phase diagram in the $\{\gamma, J^\perp\}$ plane when avalanches are defined with 3D nearest-neighbor connectivity in space (across $(x,y,k)$) and the temporal window $\Delta t_w$. Colors denote rigidity (stack-spanning avalanches), long-range order (LRO), a crossover regime, and short-range correlations. Compared to the in-plane analysis, a rigid region now exists because strong feedforward coupling merges layer-by-layer bursts into single 3D clusters. Right: Representative avalanche size distributions $P(s)$ at the starred points $\{0.1, 4.5\}$ (rigid; dominated by system-wide events), $\{0.1, 2.0\}$ (LRO; broad power-law regime), and $\{1.0, 2.0\}$ (crossover; truncated tails). Increasing $\gamma$ suppresses activity, while decreasing $J^\perp$ reduces inter-layer merging, recovering LRO-like statistics. Each data point is obtained using 1 instance simulated over 8000 time steps on a $64\times 64$ lattice.}
    \label{fig:phase_diagram_3d}
\end{figure*}

Fig.~\ref{fig:phase_diagram_3d} shows the resulting $\{\gamma,J^\perp\}$ phase diagram together with representative distributions at three starred points: $\{0.1,4.5\}$ (rigid), $\{0.1,2.0\}$ (LRO), and $\{1.0,2.0\}$ (crossover). Compared with the in–plane diagrams of Fig.~\ref{fig:phase_diagram}, the main difference is the introduction of a vast rigid region. When $J^\perp$ is large, feedforward coupling synchronizes flips across layers; many avalanches that are distinct within individual planes merge through the vertical links into a single 3D cluster that spans the stack. Consequently, power–law regimes seen in 2D are replaced by distributions dominated by system–wide events in 3D.

At intermediate couplings, two trends appear: For $J^\perp \lesssim 2$, vertical links are too weak to fully synchronize layers. Avalanches remain predominantly lateral, and the 3D distributions recover a broad algebraic regime (LRO), similar to the in–plane case. For $J^\perp > 2$ but $\gamma \sim 1$, memory weakens while vertical links are moderately strong; partial inter-layer merging also produces power-law distributions of avalanches corresponding to LRO. As $\gamma$ increases further, activity becomes sparse, producing a crossover regime which eventually gives way to short–range behavior when $\gamma$ is too large.

Taken together, the 3D clusters make explicit the mechanism behind MILRO drag: vertical coupling links together correlated planes. When $J^\perp$ is strong, this effect yields stack–spanning events (rigidity); when moderate, it transmits order without full synchronization (LRO/crossover); and when weak, layers decouple and only small clusters survive. 

\section{Absence of Forcing Between Subsequent Layers}\label{appendix:no_forcing}

From our previous results, it is still possible that the transfer of LRO arises trivially from the forcing of spins in lower layers to precisely track the dynamics of their upper counterparts, regardless of the dynamics of these lower layers. To dispel this possibility, we calculate the correlation between flipping spins in adjacent layers $k$ and $k+1$ explicitly, defined as follows:

\begin{align}
    \tilde{G}^{(k)}(\tau) = \Big\langle \text{sgn}\Big(J_i^{\perp(k)}\Big) \dot{s_i}^{(k)}(t) \dot{s_i}^{(k+1)}(t+\tau) \Big\rangle\,.\label{eq:no_forcing}
\end{align}

Above, the average is first taken over space, at all indices $i$, and then over time, at all instances $t$ when the upper layer's spin is actively flipping (i.e., its derivative is non-zero). This function explicitly depends on $\tau$, a measure of the time delay between spin flips in adjacent layers. The factor of $\text{sgn}(J^\perp)$, where $\text{sgn}(\cdot)$ returns either $+1$ or $-1$ based on the sign of its argument, is incorporated since trivial forcing corresponds to spins $s_i^{(k)}$ and $s_i^{(k+1)}$ flipping {\it oppositely} if their interaction is anti-ferromagnetic.

\begin{figure}[htbp]
    \centering
    \includegraphics[width=\columnwidth]{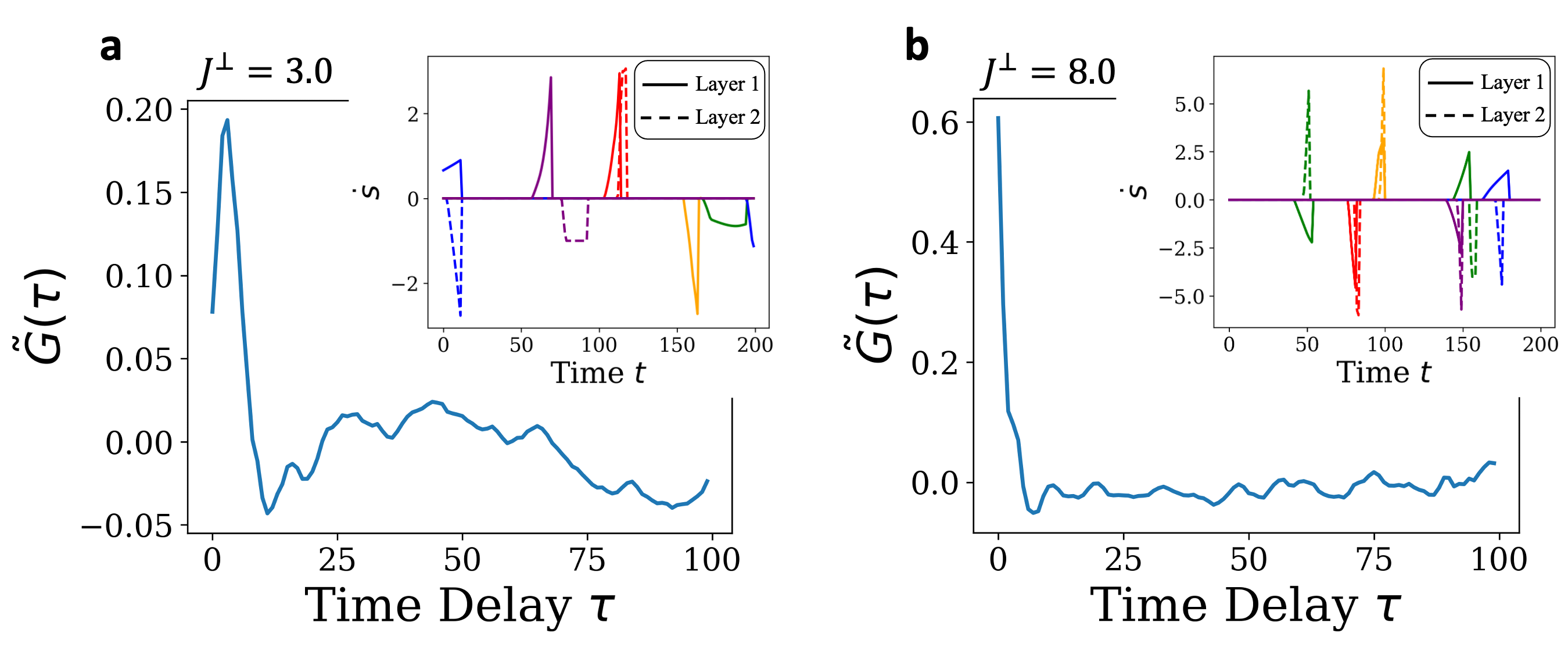}
    \caption{The correlation function $\tilde{G}^{(1)}$ as a function of $\tau$ for (a) $J^\perp = 3.0$ and (b) $J^\perp = 8.0$. In both cases, the memory frequency $\gamma=0.05$. A random sample of 5 spin trajectories are shown in the inset, with spins in the same lattice position $i$ in adjacent layers plotted in the same color. (a) $\tilde{G}$ peaks far below unity before dropping to near zero, suggesting spins in layer 2 are not being trivially forced to track the dynamics of layer 1 spins. In the inset, several layer 1 spin flips do not induce layer 2 flips at all. (b) $\tilde{G}$ peaks much closer to unity. All spin trajectories in layer 2 appear to trivially follow those in layer 1.}
    \label{fig:no_forcing}
\end{figure}

In Fig.~\ref{fig:no_forcing}, Eqn.~\ref{eq:no_forcing} is plotted as a function of time delay $\tau$ when $J^\perp = 3.0$ and $8.0$. When $J^\perp$ is not too large, $\tilde{G}$ peaks far below unity after a short time delay before dropping to approximately zero for longer delays. This suggests that these lower-layer spins do not trivially track the dynamics of the upper spins, although the directional flow of information in time can still be resolved. On the other hand, when $J^\perp = 8.0$, this peak occurs nearly instantaneously and takes a much higher value. The insets in Fig.~\ref{fig:no_forcing} showing individual spin dynamics makes the following especially clear: so long as $J^\perp$ is not so large as to dominate over intra-layer interactions, lower layers do {\it not} trivially follow the dynamics of their upper neighbors. Nevertheless, LRO can still be dragged through the layers.

\FloatBarrier

\bibliography{references}

\end{document}